\documentclass{emulateapj}
\usepackage{graphicx}
\usepackage{times}

\def\ltsima{$\; \buildrel < \over \sim \;$}
\def\simlt{\lower.5ex\hbox{\ltsima}}   
\def\gtsima{$\; \buildrel > \over \sim \;$}
\def\simgt{\lower.5ex\hbox{\gtsima}}

\def\etal{{\it et al.}}
\def\lsim{\mathrel{\lower .85ex\hbox{\rlap{$\sim$}\raise
.95ex\hbox{$<$} }}}
\def\gsim{\mathrel{\lower .80ex\hbox{\rlap{$\sim$}\raise
.90ex\hbox{$>$} }}}

\def \apjl  {ApJL}

\def\mnras {MNRAS}
\def \etal {et~al.~}
\def \chisq  {\ifmmode  \chi^2   \else  $\chi^2$  \fi}  
\def \chisqr {\ifmmode \chi^2_{\rm r} \else $\chi^2_{\rm r}$ \fi}
\def \spose#1{\hbox  to 0pt{#1\hss}}  
\def \lta{\mathrel{\spose{\lower 3pt\hbox{$\sim$}}\raise  2.0pt\hbox{$<$}}}
\def \gta{\mathrel{\spose{\lower  3pt\hbox{$\sim$}}\raise 2.0pt\hbox{$>$}}}
 
\def \ha  {\ifmmode H\alpha \else H$\alpha $ \fi}

\def \kms {\ifmmode  \,\rm km\,s^{-1} \else $\,\rm km\,s^{-1}  $ \fi }
\def \kpc {\ifmmode  {\rm kpc}  \else ${\rm  kpc}$ \fi  }  
\def \Msun {\ifmmode M_{\odot} \else $M_{\odot}$ \fi} 
\def \hMsun {\ifmmode h^{-1}\,\rm M_{\odot} \else $h^{-1}\,\rm M_{\odot}$ \fi}
\def \hhMsun {\ifmmode h^{-2}\,\rm M_{\odot}\else $h^{-2}\,\rm M_{\odot}$ \fi}
\def \Lsun {\ifmmode L_{\odot} \else $L_{\odot}$ \fi} 
\def \hhLsun {\ifmmode h^{-2}\,\rm L_{\odot} \else $h^{-2}\,\rm L_{\odot}$ \fi}

\def \LCDM {\ifmmode \Lambda{\rm CDM} \else $\Lambda{\rm CDM}$ \fi}
\def \lCDM {\ifmmode \Lambda{\rm CDM} \else $\Lambda{\rm CDM}$ \fi}
\def \lcdm {\ifmmode \Lambda{\rm CDM} \else $\Lambda{\rm CDM}$ \fi}
\def \sig8 {\ifmmode \sigma_8 \else $\sigma_8$ \fi} 
\def \OmegaM {\ifmmode \Omega_{\rm M} \else $\Omega_{\rm M}$ \fi} 
\def \OmegaL {\ifmmode \Omega_{\rm \Lambda} \else $\Omega_{\rm \Lambda}$\fi} 
\def \Deltavir {\ifmmode \Delta_{\rm vir} \else $\Delta_{\rm vir}$ \fi}

\def \rs {\ifmmode r_{\rm s} \else $r_{\rm s}$ \fi} 
\def \rrm2 {\ifmmode r_{-2} \else $r_{-2}$ \fi} 
\def \ccm2 {\ifmmode c_{-2} \else$c_{-2}$ \fi} 
\def \cvir {\ifmmode c_{\rm vir} \else $c_{\rm vir}$ \fi} 
\def \cbar {\ifmmode \overline{c} \else $\overline{c}$ \fi}

\def \R200 {\ifmmode R_{200} \else $R_{200}$ \fi} 
\def \Rvir {\ifmmode R_{\rm vir} \else $R_{\rm vir}$ \fi}

\def \v200 {\ifmmode V_{200} \else $V_{200}$ \fi} 
\def \Vvir {\ifmmode V_{\rm  vir} \else  $V_{\rm vir}$  \fi} 
\def  \Vhalo  {\ifmmode V_{\rm halo} \else $V_{\rm halo}$ \fi}

\def \M200 {\ifmmode M_{200} \else $M_{200}$ \fi} 
\def \Mvir {\ifmmode M_{\rm  vir} \else $M_{\rm  vir}$ \fi}  
\def \Mshell  {\ifmmode M_{\rm shell} \else $M_{\rm shell}$ \fi}

\def \Nvir {\ifmmode N_{\rm  vir} \else $N_{\rm  vir}$ \fi}  

\def \Jvir {\ifmmode J_{\rm vir} \else $J_{\rm vir}$ \fi} 
\def \Jshell {\ifmmode J_{\rm shell} \else $J_{\rm shell}$ \fi}

\def \Evir {\ifmmode E_{\rm vir} \else $E_{\rm vir}$ \fi} 

\def \lam {\ifmmode \lambda  \else $\lambda$ \fi} 
\def \lamp {\ifmmode \lambda^{\prime} \else $\lambda^{\prime}$  \fi} 
\def \lampc {\ifmmode \lambda^{\prime}_{\rm c} \else
  $\lambda^{\prime}_{\rm c}$  \fi} 
\def \lambar {\ifmmode \bar{\lambda}  \else  $\bar{\lambda}$  \fi}  
\def  \lampbar  {\ifmmode \bar{\lambda^{\prime}} \else
  $\bar{\lambda^{\prime}}$\fi} 
\def \siglam {\ifmmode \sigma_{\lambda} \else $\sigma_{\lambda}$ \fi} 
\def \siglamp {\ifmmode                \sigma_{\lambda^{\prime}} \else
$\sigma_{\lambda^{\prime}}$\fi}

\def \Rd {\ifmmode R_{\rm d} \else $R_{\rm d}$ \fi} 
\def \Rs {\ifmmode R_{\rm s} \else $R_{\rm s}$ \fi}  
\def \Rd {\ifmmode R_{\rm d} \else $R_{\rm d}$ \fi}  
\def \Rcool  {\ifmmode R_{\rm  cool}  \else $R_{\rm cool}$ \fi} 
\def \RIII {\ifmmode  3.2\Rs \else $3.2\Rs$ \fi} 
\def \RII {\ifmmode 2.2\Rs \else $2.2\Rs$  \fi} 
\def \Reff {\ifmmode R_{\rm eff} \else $R_{\rm  eff}$ \fi} 
\def  \rb {\ifmmode r_{\rm b}  \else $r_{\rm b}$ \fi}

\def  \Sigmacrit   {\ifmmode  \Sigma_{\rm  crit}   
\else  $\Sigma_{\rm crit}$\fi} 
\def \Sig0 {\ifmmode \Sigma_{0} \else $\Sigma_{0}$ \fi}

\def \muI {\ifmmode \mu_{0,I} \else $\mu_{0,I}$ \fi}

\def \mgal {\ifmmode m_{\rm gal} \else $m_{\rm gal}$ \fi} 
\def \md {\ifmmode m_{\rm d} \else $m_{\rm d}$ \fi} 
\def \ms {\ifmmode m_{\rm   s}   \else   $m_{\rm   s}$   \fi}   
\def   \mdbar   {\ifmmode {\overline{m}}_{\rm d} \else
  ${\overline{m}}_{\rm d}$ \fi} 
\def \msbar {\ifmmode  \bar{m}_{\rm  s}  \else  $\bar{m}_{\rm s}$
  \fi}  
\def  \Md {\ifmmode M_{\rm d}  \else $M_{\rm d}$ \fi} 
\def  \Ms {\ifmmode M_{\rm s} \else $M_{\rm  s}$ \fi} 
\def \Mb {\ifmmode  M_{\rm b} \else $M_{\rm b}$ \fi} 
\def \Mstar {\ifmmode  M_{\rm star} \else $M_{\rm star}$ \fi}
\def \Mdisc {\ifmmode M_{\rm disc} \else $M_{\rm disc}$ \fi}

\def \Jd {\ifmmode J_{\rm d} \else $J_{\rm d}$ \fi} 
\def \Jb {\ifmmode J_{\rm b} \else $J_{\rm b}$ \fi}  
\def \fb {\ifmmode  f_{\rm b} \else $f_{\rm b}$ \fi}

\def  \jd  {\ifmmode j_{\rm  d}  \else  $j_{\rm  d}$ \fi}  
\def  \jdmd {\ifmmode \frac{j_{\rm  d}}{m_{\rm d}} \else
  $\frac{j_{\rm d}}{m_{\rm d}}$ \fi} 
\def \fj {\ifmmode f_{\rm j} \else $f_{\rm j}$ \fi} 
\def \ft {\ifmmode f_{\rm t}  \else $f_{\rm t}$ \fi} 
\def  \fM {\ifmmode f_{\rm M} \else $f_{\rm M}$ \fi}

\def  \Vd {\ifmmode  V_{\rm  d}  \else $V_{\rm  d}$  \fi} 
\def  \Vcool {\ifmmode V_{\rm cool} \else $V_{\rm cool}$ \fi} 
\def \Vcirc {\ifmmode V_{\rm circ}  \else $V_{\rm circ}$  \fi} 
\def \VIII  {\ifmmode V_{3.2} \else $V_{3.2}$ \fi} 
\def  \VII {\ifmmode V_{2.2} \else $V_{2.2}$ \fi}
\def \Vobs {\ifmmode V_{\rm obs}  \else $V_{\rm obs}$ \fi} 
\def \Vdisc {\ifmmode V_{\rm disc} \else  $V_{\rm disc}$ \fi} 
\def \Vmax {\ifmmode V_{\rm  max} \else  $V_{\rm max}$  \fi} 
\def  \Vmaxobs{\ifmmode V_{\rm max}^{\rm obs}\else  $V_{\rm max}^{\rm
    obs}$\fi}  
\def \Vtot {\ifmmode V_{\rm tot} \else $V_{\rm tot}$  \fi} 
\def \Vrot {\ifmmode V_{\rm rot} \else  $V_{\rm rot}$  \fi} 
\def  \Vflat {\ifmmode  V_{\rm  flat} \else $V_{\rm flat}$ \fi}

\def \Ups {\ifmmode \Upsilon  \else $\Upsilon$ \fi} 
\def \YB {\ifmmode \Upsilon_B \else $\Upsilon_B$ \fi} 
\def \YI {\ifmmode  \Upsilon_I  \else $\Upsilon_I$ \fi} 
\def \DeltaIMF {\ifmmode \Delta_{\rm IMF} \else $\Delta_{\rm IMF}$ \fi}

\def\LCDM{$\Lambda$CDM }

\def\c200{$c_{200}$}

\shorttitle{}
\shortauthors{}

\begin{document}

\title{Breaking up the Magellanic Group into the Milky Way Halo: understanding the local dwarf galaxy properties}

\author{Elena D'Onghia\altaffilmark{1,2}\\ }
\affil{Institute for Theoretical Physik, 
 University of Zurich, Winterthurerstrasse 190, 8057 Zurich, Switzerland}
\email{elena@physik.unizh.ch}


\altaffiltext{2}{Marie Curie Fellow}

\begin{abstract}
We use a numerical simulation of a loose group containing a  Milky Way halo to probe that in the hierarchical universe  
the Magellanic Clouds and some dSphs have been accreted into the Milky Way halo from a late infalling group of dwarfs.
Our simulations show  that  the tidal breakup of the Magellanic group occurs before it enters the Milky Way halo. 
Only half of the satellites contributed from the group are predicted to be inside  the Milky Way virial radius. 
Half of its subhalos  survive outside the current virial radius in the form
of satellites, whereas the remaining material contributes to the diffuse Milky Way halo.
At $z\sim 0$ the disrupted group contributes less than 10\% to the 
Milky Way halo mass but 20\% of the  brightest dwarf galaxies of the Milky Way   have been part of this group.
This scenario  points out that some dSphs might have been formed away from giant spirals and been accreted already as  spheroids,
by a late infall group in contrast with the classical picture of tidal stripping of dSph formation models. 
This would naturally 
explain several peculiarities of the local dSph: why Draco and the other luminous dSphs exist compared to other ultra-faint satellite galaxies,  
the location of Tucana and Cetus in the outskirts 
of the Local Group and the mismatch in metallicity between the stellar halo of the Milky Way and the dwarf galaxies that many have suspected dissolved to build it.

\end{abstract}


\section{Introduction}
An unequivocal  prediction of  cold dark matter models (CDM) is that the mass of the Milky Way, in the form of its dark matter halo,
builds up hierarchically, by accretion of lower-mass halos. When these sub-systems (often refereed to as subhalos) 
escape the tidal disruption in the Milky Way halo
they survive in the form of satellite galaxies until today. Numerical simulations confirmed the theory predicting 500 satellites within 500 kpc
from the Milky Way center (Kauffman et al. 1993; Klypin et al. 1999; Moore et al. 1999). 
The modest populations of observed dwarf galaxies orbiting the Milky Way and Andromeda, however, seems to conflict with this prediction. 
Indeed, in the current CDM paradigm of structure formation, dark matter satellites 
outnumber the known spheroidal by a factor of 10 to 100. 
This discrepancy between the expected and observed numbers of dwarf galaxies 
has become widely known as the {\it missing dwarf problem}. The newly discovered population of ultra-faint dwarfs around the Milky Way and M31
found in the Sloan Digital Sky Survey (Willman et al. 2005a,b; Belokurov et al. 2006, Zucker et al. 2006)  
increases of a factor of two in the number of known 
satellites. It is unclear, however, whether these new discovered low surface brightness satellite galaxies can 
reconcile the outnumber of subhalos predicted in cosmological simulations and solve the problem (Simon \& Geha 2007).

So far a number of mechanisms have been proposed 
to the substructure problem. 
Cosmological solutions include modifying the power spectrum at small scales (Kamionkowski \& Liddle 2000; Zentner \& Bullock 2003) and changing the 
nature of the dark matter particles, such as assuming a warm dark matter particle (e.g. Colin et al. 2000; Avila-Reese et al. 2001) 
or invoking a decay from a 
nonrelativistic particle (Strigari et al. 2007). Alternative solutions typically appeal to feedback effects associated with stellar evolution or including 
heating from UV radiation in order to  suppress the formation of dwarf galaxies by preventing 
low mass dark matter halos from acquiring enough gas to form stars (e.g., Bullock et al. 2000; Somerville 2002; Benson et al. 2002).
An additional scenario has been proposed by Kravtsov, Gnedin \& Klypin (2004) suggesting  that the dwarf spheroidal 
we observe today were once much more massive 
objects that have been reduced to their present mass by tidal stripping processes. 

Although these models can reconcile the shallow faint end of the luminosity function of galaxies with the low-
mass end of the halo mass function, they did not explain so far the peculiarities of the dwarf spheroidal 
of the Local Group.
For instance, it is known that dSphs of the Local Group tend to cluster tightly around the giant spirals.
Proximity to a giant central galaxy prevent them from accreting material 
and from further star formation and allow for tidal interactions to convert them from dwarf irregulars
into spheroids. However, the presence of isolated dSphs found in the outskirts of the Local Group like Tucana or Cetus (Grebel, Gallagher \& Harbeck 2003)
seem to imply that dSphs might have been formed away from giant spirals and been accreted already as  spheroids,
in disagreement with the tidal stripping scenario, which is the common picture of dSph formation models (see e.g. Mayer et al. 2007 and reference therein).

Clues to these questions may be gathered from the current observations of the
metallicities of a large sample of stars in four nearby dwarf
spheroidal galaxies: Sculptor, Sextans, Fornax, and Carina. Recent work from Helmi et al. (2006) shows that all the four systems  
lack significantly of stars with low metallicity
with the dSph metallicity distribution significantly different from that of the Galactic halo. 
In this context Sales et al. (2007) pointed out that satellite galaxies might be accreted into the main halo as 
multiple systems
and more recently Li \& Helmi (2007)  showed that subhalos are often accreted in small groups.
Lake \& D'Onghia (2008) have collected data for nearby dwarf associations from Tully et al. (2006) including the list of candidates of 
dwarf galaxies associated with the Magellanic plane group: LMC, SMC, Sagitarius, Ursa Minor, 
Draco, Sextans and LeoII (Lynden Bell 1976, Fusi Pecci et al. 1995; Kroupa \etal 2005). The authors showed that 
seven out of ten dwarfs within $\sim 200 $kpc from the Milky Way center  might well be part of a group of 
dwarfs recently accreted into the Milky Way showing that there are natural mechanisms 
that lead to less suppression of satellite galaxies in dwarf galaxy groups and pointing to a resolution  for the 
missing dwarf problem in the Milky Way. 

In this paper, we present a simulation of a loose group containing a Milky Way sized halo showing that the Magellanic group 
likely entered the Milky Way  at a redshift 1 and breaks up outside the virial radius. In Lake \& D'Onghia (2008) we showed that
this picture can naturally explain why galaxy satellites like Draco or Ursa Minor are luminous and why other dwarfs are expected to be dark.  
Here we focus on the peculiarities of the satellite galaxies of the Local Group and show that they are consistent with the tidal 
breakup of an infalling group 
of dwarf galaxies dominated from the LMC and SMC systems.

\section{NUMERICAL METHODS}
\subsection{The Numerical simulation}
Our analysis is based on a high-resolution cosmological simulation of a loose group containing a Milky Way sized halo.
The target halo of the loose group has a virial mass $\sim$7x$10^{12} \ h^{-1}$M$_{\odot}$ with $\sim$ 6 millions particles 
within the virial radius. The group 
was identified in a cosmological simulation 
of box 90 Mpc (or 67.5 $h^{-1}$ Mpc) (comoving) on a side
of 300$^3$ particles with cosmological parameters chosen to match the 3-yr Wilkinson Microwave Anisotropy Probe (WMAP3) constraints (Spergel et al. 2006).
These are 
characterized by the
present-day matter density parameter, $\Omega_0=0.238$; a cosmological
constant contribution, $\Omega_{\Lambda}$=0.762; and a Hubble
parameter $h=0.73$ ($H_0=100\, h$ km s$^{-1}$ Mpc$^{-1}$). The mass
perturbation spectrum has a spectral index, $n=0.951$, and is
normalized by the linear rms fluctuation on $8 \ h^{-1}$ Mpc radius
spheres, $\sigma_8=0.75$.  The candidate halo  was selected according to the following criteria:
{\it (i)} Environment location along a filament at least 5 $h^{-1}$ Mpc  far from any rich galaxy group or galaxy cluster at z=0.
{\it (ii)}  The the peak velocity of the Milky Way halo inside the group is 206 km s$^{-1}$ 

The group candidate containing the Milky Way halo was selected and resimulated to higher resolution using 
GRAFIC2 (Bertschinger 2001).  
\begin{figure}
\begin{center}
\includegraphics[width=83mm]{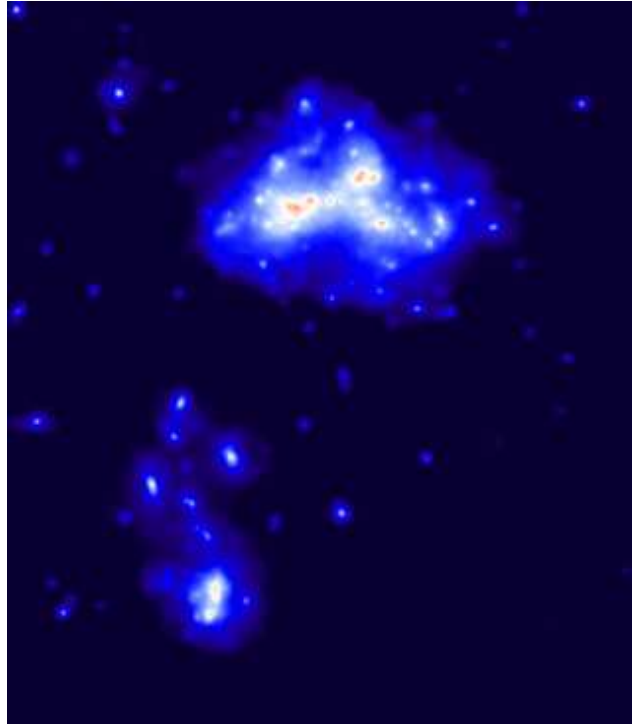}
\end{center}
\caption{The Magellanic group approaching the Milky way halo}
\end{figure}
The simulation was run with the tree-code PKDGRAV (Stadel 2001). Gravitational interactions 
between pairs of particles are softened with a fixed comoving softening length of 0.8 $h^{-1}$ kpc.

\subsection{The Analysis}
We use SKID (Stadel 2001) to identify subhalos  in the high resolution region. SKID finds subhalos within main halos
by locating high density regions within the main halo and identifying the bound group of particles associated with each overdensity
using a friends-of-friends (FOF) method. Particles which are not bound to their group are removed from the group by an unbinding procedure.
We list for our catalog all subhalos with more than 20 particles. The output of SKID is a list of subhalos with their structural properties.
We focus on the following satellite properties: {\it( i)} the subhalo mass M$_{\rm{sat}}$; {\it (ii)} the peak of the subhalo circular velocity profile 
$v_{\rm{cir}}$; {\it (iii)} the location of the subhalo center, identified with the most bound particle, which is the particle 
with the minimum gravitational potential energy.
SKID was run on all snapshot up to z=5, allowing us to track the evolution of individual subhalos.

\begin{deluxetable}{rrrr}
\tablecolumns{4}
\tablewidth{0pc}
\tablecaption{Subhalos of the disrupted group at z=0 }
\tablehead{
\colhead{N$_{\rm{subhalos}}$} & \colhead{V$_{\rm{max}}$ [km s$^{-1}$]} &  \colhead{M$_{\rm{sat}}$ [M$_{\odot}$]} & \colhead{Distance [kpc]}
}
\startdata
\tableline
1   &  62   &  1.2x10$^{10}$  & 134 \\
1   &  53   &  6.5x10$^{9}$   & 189\\
1   &  40   &  4x10$^{9}$     & 720\\
1   &  35   &  2.5x10$^{9}$   & 20\\
2   &  29   &  2x10$^{9}$     & 69-189\\
1   &  27   &  1x10$^{9}$     & 1072\\
1   &  18   &  4.5x10$^{8}$   & 341\\
2   &  16   &  4x10$^{8}$     & 55-770\\
1   &  14   &  2.9x10$^{8}$   & 277\\
3   &  13   &  3x10$^{8}$     & 153-247-352\\
3   &  12   &  1.5x10$^{8}$   & 58-66-131\\
1   &  11   &  1.3x10$^{8}$   & 85\\
5   &  10   &  1.3x10$^{8}$   & 39-45-344-404-734\\
\enddata
\end{deluxetable}

\section{RESULTS}
\subsection{The Infalling Magellanic Group  into the  Milky Way}
We have identified all the surviving subhalos within the virial radius of the Milky Way halo at present time.
Within the virial radius of the simulated Milky Way, there are a total of 70 satellites with circular velocities greater than 10$\kms$ at present time.  
There are five Magellanic Clouds-sized subhalos inside the Milky Way halo with peak circular velocities between 50 and 62 km s$^{-1}$.
We traced each representative
subhalos backwards in time and we searched for the group it belongs to at preceding snapshot.
We focused on two target subhalos with circular velocity of  62 and 53 $\kms$ respectively and found out that these two subhalos
were both satellites of an infalling group of substructures.  Hereafter we refer to this infalling group as the ``Magellanic group''.   

Figure 1 shows the Magellanic group (at the bottom the picture) infalling into the Milky Way halo at z=1.12. 
The group infalls late and it breaks up outside the virial radius of the final Milky Way halo and spreads its surviving satellites  over such a wide range of radii.
The infalling group of satellites is loose and resembles association of 
dwarf galaxies discovered from Tully et al. (2006) within 1 and 3 Mpc from the Local Group.
Surprisingly  the group becomes unbound through tidal disruption before entering within the Milky Way halo.

\begin{figure}
\begin{center}
\includegraphics[width=83mm]{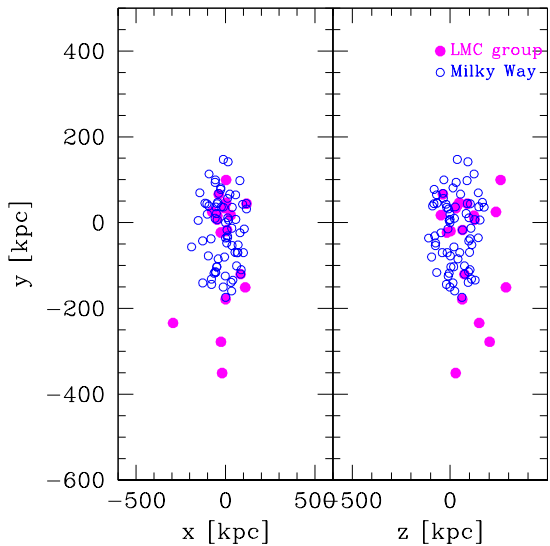}
\end{center}
\caption{The spatial distribution of the satellites within the virial radius of the Milky Way (blue open circles) as compared
to the contributed subhalos of the break up Magellanic  group at z=0 within 600 Kpc from the Milky Way center
(magenta filled circles). The remaining subhalos of the disrupted group in the outskirts of the loose group are listed in Table 1.}
\end{figure}

The basic properties of the members of the Magellanic group at z=0 are presented in Table 1. The group has roughly only 23 members surviving as 
satellites 
which contribute to the total budget of subhalos of the Milky Way at present time.
Table 1 lists the total number of subhalos found with similar peak of the circular velocity, $v_{\rm{max}}$ and the mass associated to the members.

Figure 2 shows the spatial distribution of all the satellites within the virial radius of the Milky Way (blue filled circles) as compared
to the subhalos of the disrupted Magellanic group at z=0  (magenta stars).  Despite the late infall, this particular group appears very well mixed, however
almost half of the surviving  subhalos of the group (10 subhalos) are at present time located outside the virial radius of the final Milky Way.  
A few of them are in the outskirts  of the Milky Way. These subhalos may intuitively reproduce the special cases like Tucana or Cetus which seem to be located 
in the outskirts of the Local Group (Grebel, Gallagher \& Harbeck 2003). 
In particular one of the satellite of the destroyed group is almost 1 Mpc away from the Milky Way center (see table 1).
Out of 23 surviving subhalos of the LMC group the remaining subhalos and material of the group  is destroyed and contributes to the diffuse Milky Way halo.
We traced backwards in time  the other representative Magellanic systems and found that they belong to infalling multiple systems. 
Figure 3 shows the peak circular velocity cumulative distribution of the satellites contributed by the simulated infalling Magellanic group 
measured at z=0 compared to the observed Magellanic plane group (filled squared symbols)(see also Figure 2 in Lake \& D'Onghia 2008).
The  Magellanic plane group includes: LMC, SMC, Sagittarius, Ursa Minor, Draco, Sextans and Leo II (Lynden Bell 1976; Fusi Pecci et al. 1995; Kroupa \etal 2005).
The Figure displays also the circular velocity cumulative distribution of the observed satellites in the Milky Way including
the newest dwarfs
with a minimum $\sigma = 3.3\kms$ and a correction for incomplete sky coverage (Simon \& Geha 2007).
We note that  while the simulated Magellanic group contributes less than 10\% of the Milky Way halo mass, it is responsible for nearly 20\% of the
subhalos within the Milky Way virial radius at present time.    
Having only 10\% of the Milky Way mass any accreted gas inside the group would cool down quicly preventing its dwarf members from
being stripped by ram pressure. As a remarkable consequence the dSphs might have been  accreted into the Milky Way already as spheroids. 
\begin{figure}
\begin{center}
\includegraphics[width=83mm]{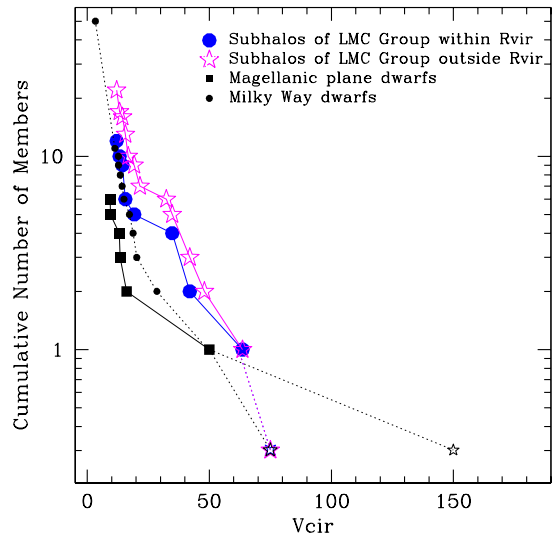}
\end{center}
\caption{ The peak circular velocity cumulative distribution of the satellites contributed by the simulated infalling Magellanic group 
measured at z=0 inside the virial radius of the Milky Way halo (blue filled points ) and outside the virial radius (open magenta symbols).
The velocity cumulative distribution of the observed  Magellanic group including LMC, SMC, Sagittarius, Ursa Minor, Draco, Sextans and Leo I
is displayed with black filled squared. The little filled circles shows the velocity cumulative distribution of satellite galaxies
in the Milky Way  inferred from Simon \& Geha (2007). }
\end{figure}
\begin{figure*}
\begin{center}
\includegraphics[width=166mm]{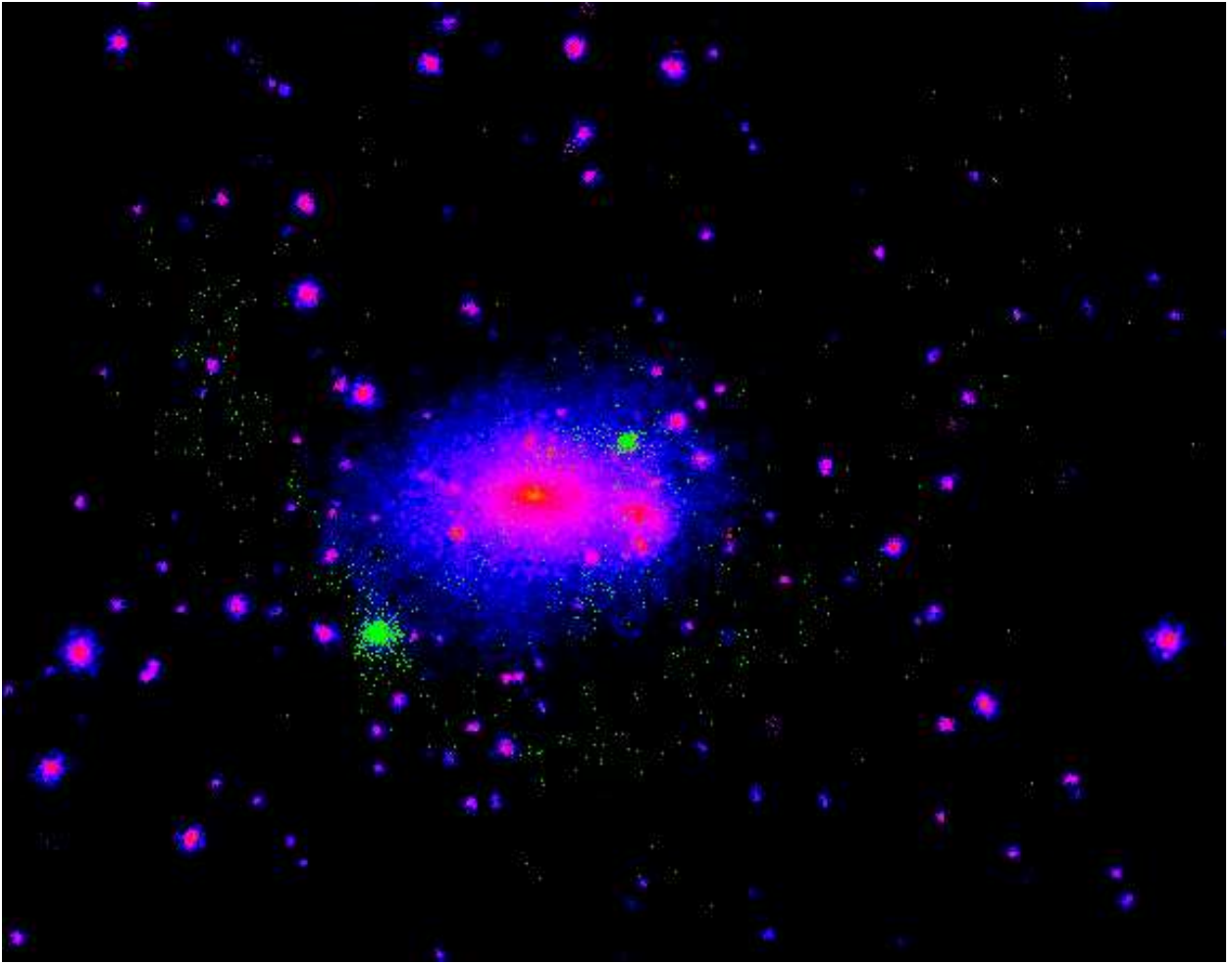}
\end{center}
\caption{The two target subhalos of 62 and 53 $\kms$ representative in our sample of the Magellanic Clouds (marked with green color) 
orbiting within the Milky Way at z=0.2. Elongated tidal streams extend outside the virial radius (streams are marked with green color). }
\end{figure*}
Figure 4 shows the two target subhalos of 62 and 53 $\kms$ representative in our sample of the Magellanic Clouds orbiting within the Milky Way halo.
Elongated tidal streams are formed well outside the virial radius. These two effects, the survivor satellites and long tidal debris outside
the virial radius reinforce recent findings suggesting that the traditional virial radius and virial mass of the dark halos might be underestimated 
of a factor of two or three when a bulk infall is occurring (e.g. see Figure 3 in D'Onghia \& Navarro 2007; Cuesta \etal 2007).
This is also consistent with results from N-body simulations suggesting that many satellite galaxies are bound on extreme orbits with apocentric
distances of three virial radii (e.g. Moore et al. 2004; Diemand et al. 2007; Sales et al. 2007).

\section{DISCUSSION AND CONCLUSION}
We performed a simulation of a Milky Way halo in the $\Lambda$CDM models focusing on the properties of the satellite population at the present day.
Our simulation probes that in the cosmological context of the hierarchical universe  
the Magellanic Clouds might have been accreted into the Milky Way from an infalling group and that the LMC and SMC became unbound only recently. 
This results is consistent with the recent findings from Kallivayalil et al. (2006), Besla et al. (2007)   
who argue that the LMC fell into the Milky Way halo only 2 Gyrs ago and is moving at nearly the escape velocity at its radius approaching its orbital pericenter for the first time.
One might argue that the time of the accretion of Magellanic Clouds is $z\sim 1$ in this specific simulation, much earlier than the observational results
but groups with 10 \% of the Milky Way mass may infall any time 

Our main results may be summarized as follows. {\it (i)} The tidal break up of the Magellanic group occurs outside the current virial radius of the Milky Way 
and spreads its surviving satellites  over 2 or 3 virial radii. Half of its satellites survive outside the current virial radius of the Milky Way in the form
of satellites, whereas the remaining material contributes to the diffuse Milky Way halo. {\it (ii)} At z=0 the disrupted Magellanic group contributes less than 10\% to the 
Milky Way mass but 20\% of the  brightest dwarf galaxies of the Milky Way   have been part of this group.
{\it (iii)} The circular velocity cumulative distribution of the satellites
of the simulated Magellanic group matches the function of the observed Magellanic plane satellites   
the in the Local Group, suggesting that the gas physics in dwarf groups plays a key role in keeping dwarf galaxies luminous and might be
the key to solve the missing dwarf problem
as already discussed in Lake \& D'Onghia (2008). 
{\it (iv)} Elongated tidal streams are formed as remnants of the breakup of the Magellanic group 
well outside the traditional virial radius estimate, reinforcing the idea that the traditional estimates of the virial masses around galaxies might be underestimated
of a factor of 2 when an infall is happening.
 A remarkable consequence of the last result is that our models predict a great amount of low mass satellites to be discovered well
outside the virial radius of the Milky Way. It predicts groups of dwarf galaxies to be in the neighborhood of the Local Group.
Some associations of dwarf galaxies within 5 Mpc from the Milky Way have been already discovered (Tully et al. 2006)
but more associations are expected in the $\Lambda$CDM models and might be a challenge for future observations.
In particular these association of dwarfs have the properties expected of bound systems with 1-10x10$^{11}$ M$_{\odot}$, but
they have too little gas and too few stars with a consequent mass-to-light ratio of 100-1000 M$_{\odot}$/L$_{\odot}$.
Such a high  mass-to-light ratio breaks the relationship between the  M$_{\odot}$/L$_{\odot}$
and the mass of bound systems and leads to a sharp increase of  M$_{\odot}$/L$_{\odot}$ for systems below the mass of 6x10$^{11}$ M$_{\odot}$.
A similar break and sharp increase of the  M$_{\odot}$/L$_{\odot}$ at low mass systems has been pointed out from
van den Bosch et al. (2003; 2005). The authors found a paucity of light at low mass systems comparing the shallow 
faint-end of the observed luminosity functions of galaxies  with  the corresponding halo mass function.

It has not escaped our notice that the scenario 
we proved explains a lot of properties and peculiarities of the Milky Way
substructures. It suggests that dwarf galaxies have been  form in a different group environment away from the giant spirals
and been accreted  already as spheroids. 
This would naturally explain
the mismatch in metallicity between the stellar halo of the Milky Way
and the dwarf spheroidals that many have suspected dissolved to build it.

\acknowledgments
The simulations were run on zbox2 at University of Zurich. The author thanks George Lake for useful discussions, Joachim Stadel
for providing the tree-code PKDGRAV and Doug Potter and Peter Engelmaier for technical support.
She is grateful for the hospitality at the  Harvard-Smithsonian Center for Astrophysics. In particular she thanks 
Lars Hernquist, Gurtina Besla, Nitya Kallivayalil, Beth Willman and all the theory group for many useful discussions during the theory meetings
which have improved this work. This work has been supported by a EU Marie Curie fellowship under contract MEIF-041569.



\end{document}